\pdfoutput=1

\documentclass{appolb}

\usepackage{amssymb}
\usepackage{hyperref}
\usepackage{graphicx}
\usepackage{amsmath}
\usepackage[numbers,sectionbib,sort&compress]{natbib}

\begin{document}

\title{Excited Hadrons and Quark-Hadron Duality%
\thanks{Talk presented by ERA at Excited QCD 2017, 
7-13 May 2017, Sintra, Portugal}
\thanks{Supported by Spanish Mineco Grant FIS2014-59386-P, 
and by Junta de Andaluc\'{\i}a grant FQM225-05 (ERA), 
CICYTFEDER-FPA2014-55613-P,
2014-SGR-1450 and the CERCA Program/Generalitat de
Catalunya (PM).}}
\author{E. Ruiz Arriola$^1$\thanks{earriola@ugr.es}, 
P. Masjuan$^{2}$\thanks{masjuan@ifae.es} and 
W. Broniowski$^{3,4}$\thanks{Wojciech.Broniowski@ifj.edu.pl} 
\address{$^1$Departamento de F\'isica At\'omica, Molecular y Nuclear and \\ Instituto Carlos I de Fisica Te\'orica y Computacional,  Universidad de Granada, E-18071 Granada, Spain}
\address{${}^{2}$ Grup de F\'{\i}sica Te\`orica, Departament de
  F\'{\i}sica, Universitat Aut\`onoma de Barcelona  and Institut de
  F\'{\i}sica d'Altes Energies (IFAE), The Barcelona Institute of
  Science and Technology (BIST), Campus UAB,\\ E-08193 Bellaterra
  (Barcelona), Spain}
\address{$^{3}$Institute of Physics, Jan Kochanowski University, PL-25406~Kielce, Poland}
\address{$^{4}$The H. Niewodnicza\'nski Institute of Nuclear Physics, \\ Polish Academy of Sciences, PL-31342~Cracow, Poland}}

\maketitle

\begin{abstract}
We review how Quark-Hadron Duality (QHD) for (u,d) flavors at high
energies and in the scaling regime suggests a radial and angular
behaviour of mesonic and baryonic resonance masses of the Regge form
$M^2_{nJ} = \mu^2 n + \beta^2 J + M_0^2 $. The radial mass dependence
is asymptotically consistent with a common two-body dynamics for
mesons and baryons in terms of the quark-antiquark ($q \bar q)$ and
quark-diquark ($qD$) degrees of freedom, respectively.  This formula
is validated phenomenologically within an uncertainty determined by
half the width of the resonances, $\Delta M_{nJ}^2 \sim \Gamma_{nJ}
M_{nJ}$. With this error prescription we find from the non-strange PDG
hadrons different radial slopes $\mu^2_{\bar qq}=1.34(4) {\rm GeV}^2$
and $\mu^2_{qD}=0.75(3) {\rm GeV}^2$, but similar angular slopes
$\beta_{\bar q q}^2 \sim \beta_{ q D}^2 \sim 1.15 {\rm GeV}^2$.
\end{abstract}


%\bigskip \bigskip

\section{Introduction}

%Quantum Chromo-Dynamics (QCD) is the fundamental non-abelian gauge
%theory of strong interactions in terms of $ 2 N_c N_f $ quarks and
%antiquarks and and $N_c^2-1$ gluons with $N_c=3$ the number of colours
%and $N_f=6 $ the number of flavours species $u,d,s,c,b,t$. 
Confinement requires hadronic physical states to be colour singlet, but 
what are
the complete set of eigenstates of QCD spanning the Hilbert space
${\cal H}_{\rm QCD}$?.  
In the hadronic sector with light (u,d)
quarks, besides the normalizable bound states such as $\pi^+$,
$\pi^-$, $\pi^0$ or $n$, $\bar n$ and $p$, $\bar p$ (and, of course,
stable atomic nuclei and anti-nuclei, such as $^2$H,$^3$H,$^3$He, $^4$He, etc.), all
other states occur in the continuum as
asymptotic states. 
Most of the states reported by the
Particle Data Group (PDG)~\cite{Olive:2016xmw} are not bound states
but unstable resonances such as $\sigma$, $\rho$, $\omega$, $a_1$ or $\Delta$, $N^*$, 
which in a pure $(u,d)$ world would be produced from and would decay
into pions and nucleons, subject to the selection rules imposed by
conservation laws. So far, the states fitting into the quark model
classification enter the PDG tables, so this is a practical definition of
completeness, namely $ {\cal H}_{\rm PDG} = {\cal H}_{\bar q q} \oplus
{\cal H}_{qqq} \oplus {\cal H}_{\bar q \bar q \bar q} \oplus \dots$. 
However, in the case of baryons more states have been theoretically
predicted than experimentally found, hence these {\it missing
resonances} defy this criterium.  In a finite box, such as in lattice
QCD, due to the quark and gluon field boundary conditions, all states
are normalizable and their energies are discretized, hence
resonances are associated {\it only} with those states whose energies
are insensitive to the volume of the box, such that indeed ${\cal
H}_{\rm PDG} \subset {\cal H}_{\rm QCD}$.

This talk is based on Refs.~\cite{Masjuan:2012gc,Masjuan:2017fzu}
where we point out that, at least asymptotically, quark-hadron duality
(QHD) (for a review see e.g.Ref.~\cite{Melnitchouk:2005zr}) in
inclusive processes and in the scaling regime, i.e, for energies much
larger than the resonance widths $\sqrt{s} \gg \Gamma$, sets a limit
on the hadronic squared mass density for states with fixed quantum
numbers $J^{PC}$.
%in terms of radial Regge trayectories where $M_{nJ}^2 \sim \mu_J^2 n $
%characterizes an effective two-body dynamics with a linear long
%distance potential.

\section{Quark-Hadron duality}

\begin{figure}[t]
\begin{center}
\includegraphics[width=0.9\textwidth]{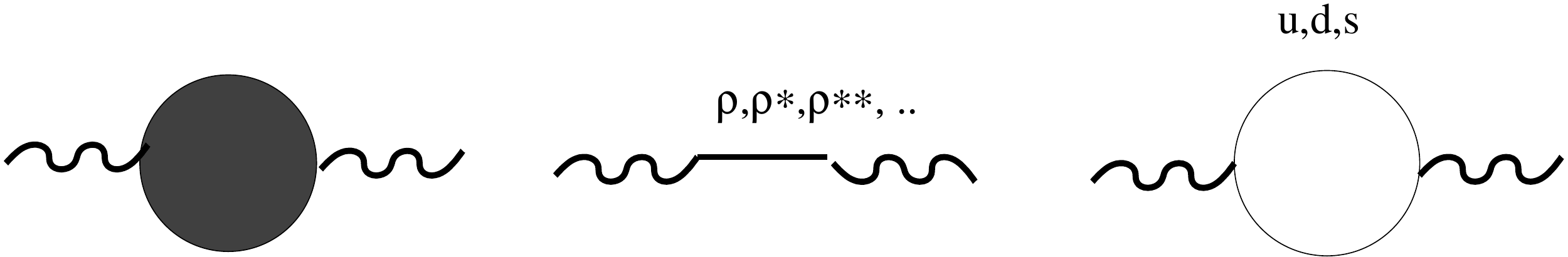}
\end{center}
\vspace{-3mm}
\caption{Quark hadron-duality for VV correlators on the vacuum
\label{fig:duality}}
\end{figure}

The meaning of QHD can be best illustrated with a simple
case. Let us consider the (conserved) vector  current $B_\mu = \bar
q \gamma_\mu q$ which vanishes in the vacuum, $\langle B_\mu \rangle
=0, $ and compute the correlator represented by Fig.~\ref{fig:duality}
(left), 
\begin{eqnarray}
\Pi_{\mu \nu}(q) = \int d^4x e^{iq\cdot x}\,
i \Big\langle 0 \Big| T \left\{ B_{\mu}(x) B_{\nu} (0) \right\}
\Big| 0\Big\rangle\, = \left( -g^{\mu\nu} + \frac{q^\mu q^\nu}{q^2}\right) \Pi(q) \, . 
\label{eq:corr}
\end{eqnarray}
At the hadronic level we assume a complete set of states, see
Fig.~\ref{fig:duality} (middle), characterized by Proca vector fields
(such as $\omega$, $\omega'$, $\omega''$, \dots), as stable and elementary
particles with masses $M_n$ and vacuum decay amplitudes 
$\langle 0 | B^\mu (0) | \omega_n^\nu \rangle = f_n q^\nu \epsilon^\mu$, with $\epsilon \cdot q=0$ implying gauge invariance
$q_\mu \Pi^{\mu\nu}=0$.  For $s=q^2 \to \infty$ we replace the sum
over $n$ by an integral and get
\begin{figure}[b]
\begin{center}
\includegraphics[width=0.9\textwidth]{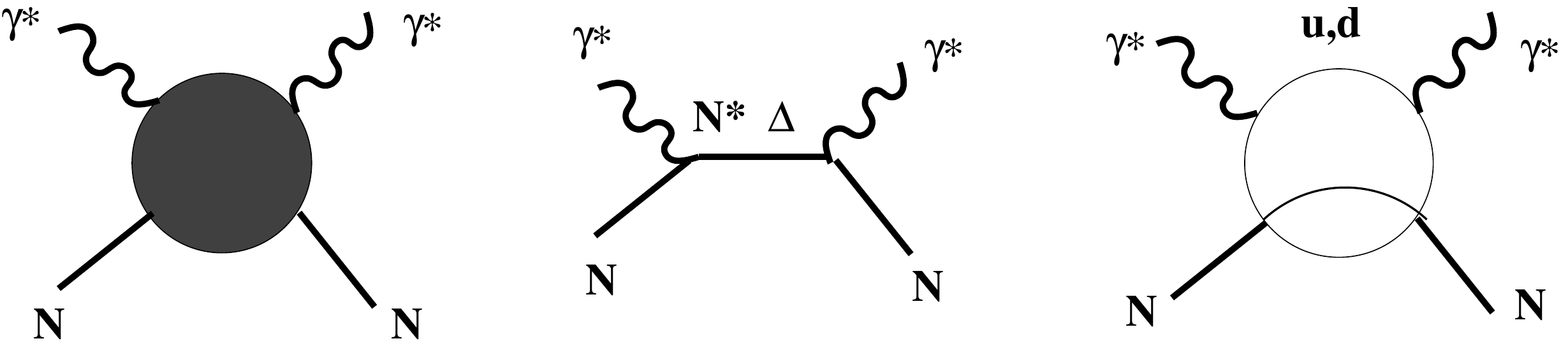} 
\end{center}
\vspace{-3mm}
\caption{Quark hadron-duality for VV correlators on the nucleon 
\label{fig:dualityN}}
\end{figure} 
\begin{eqnarray}
\frac1{\pi}{\rm Im} \Pi(s) =  \sum_{n=0}^\infty f_n^2 \delta(s-M_n^2) \to \lim_{n \to \infty}\frac{f_n^2}{d M_n^2 /d n} \, , 
\label{eq:dirac}
\end{eqnarray}
for the absorptive part.  At the quark level we have a loop with
baryon charge of the quarks equal to $1/N_c$, see
Fig.~\ref{fig:duality} (right), and massless quarks, with ${\rm Im} \Pi(s) \to 1/(24\pi N_c)$ implying
%\begin{eqnarray}
%\Pi (s) \to \sum_{i=1}^{N_c} (1/N_c)^2 \log (-s) \qquad \to
%\frac{1}{\pi}{\rm Im} \Pi(s) \to \frac{1}{N_c} \, .  
%\end{eqnarray}
the asymptotic condition encoding QHD
\begin{eqnarray}
\lim_{n \to \infty} f_n^2 /(d M_n^2 /d n)  = 1/( 24\pi^2 N_c) \, . 
\label{eq:qhd}
\end{eqnarray}
A similar discussion can be conducted for excited baryons when the
forward scattering amplitude for the nucleon of Fig.~\ref{fig:dualityN}
(left) is considered~\cite{Domokos:1971aw}, 
\begin{eqnarray} \label{eq:ncorr}
&& W_{\mu \nu} (p,q) = \frac{1}{4\pi} {\rm Im}
\int d^4x e^{iq\cdot x}\,
\sum_\lambda i \Big\langle N_ \lambda(p) \Big| T \left\{ B_{\mu}(x) B_{\nu} (0) \right\}
\Big| N_\lambda (p)\Big\rangle\, \\ 
& = & \left(-g_{\mu \nu} + \frac{q_{\mu} q_{\nu}}{q^2}\right)
W_{1} (s,q^2) +\left(p_{\mu} - q_{\mu}\frac{p\cdot q}{q^2}\right)
\left(p_{\nu} - q_{\nu}\frac{p\cdot q}{q^2}\right)
\frac{1}{M_N^2} W_{2} (s ,q^2), \nonumber 
\end{eqnarray}
with $s=(p+q)^2 $. At the hadronic level, Fig.~\ref{fig:dualityN}
(middle), we insert resonance states $|p+q, J \nu n \rangle $ with
masses $M_{J\nu n} $ and transition form factors $G_{J\nu n}^{(i)}(q)$,
to get (more details are in Refs.~\cite{Masjuan:2017fzu,Domokos:1971aw})
\begin{eqnarray}
W_i (s,q^2)= \sum_{J\nu n} [ G_{J\nu n}^{(i)}(q)]^2  \delta(s-M_{J \nu n}^2) \, .  
%W_1 (s,q^2)= \sum_{nJ\nu} \frac1{\pi}
%\frac{G_{J\nu n}(q)^2  \Gamma_{J \nu n} M_{J \nu n}}{(s-M_{J \nu n}^2)^2 + \Gamma_{J \nu n}^2 s}
\end{eqnarray}
In the Bjorken limit, $ Q^2\to\infty $ with $x=Q^2/2p\cdot q $ fixed,
and $s=M_N^2 + Q^2 (1/x-1)$.  At the quark level one obtains,
Fig.~\ref{fig:dualityN} (right), both scaling $W_1(s,q^2)\, \to
F_1(x)$, $W_2(s,q^2)\, \to F_2(x)$ and the Callan-Gross relation $
F_2(x)= 2 x F_1(x)$ due to the spin 1/2 nature of partons. The QHD
requires
\begin{eqnarray}
\sum_{J\nu}\lim_{n \to \infty} [G_{J \nu n}(q)]^2 /(d M_{J\nu n}^2 /d n)  = F_i(x) \, . 
\label{eq:qhd-bar}
\end{eqnarray}
which is satisfied if  $G_{J\nu n}(q) \to F_{J\nu}(-Q^2/M_{J\nu n}^2) $ and 
$d M_{J\nu n}^2 /d n \to \mu_{J\nu}^2$. 

\begin{figure}[b]
\begin{center}
\includegraphics[width=0.49\textwidth]{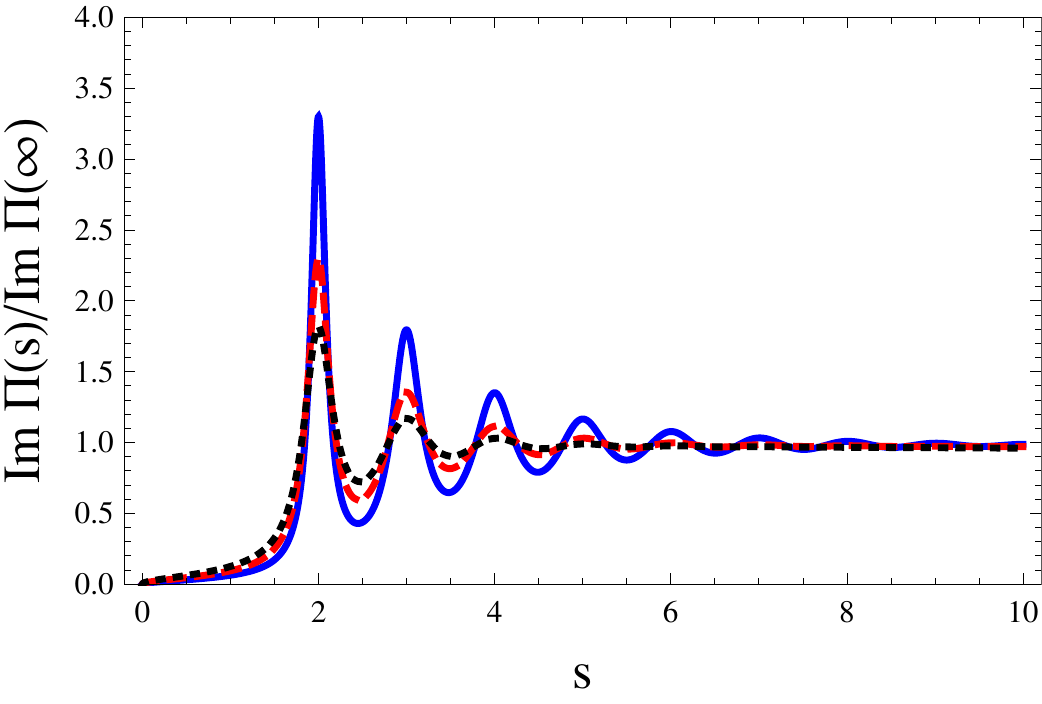}
\includegraphics[width=0.49\textwidth]{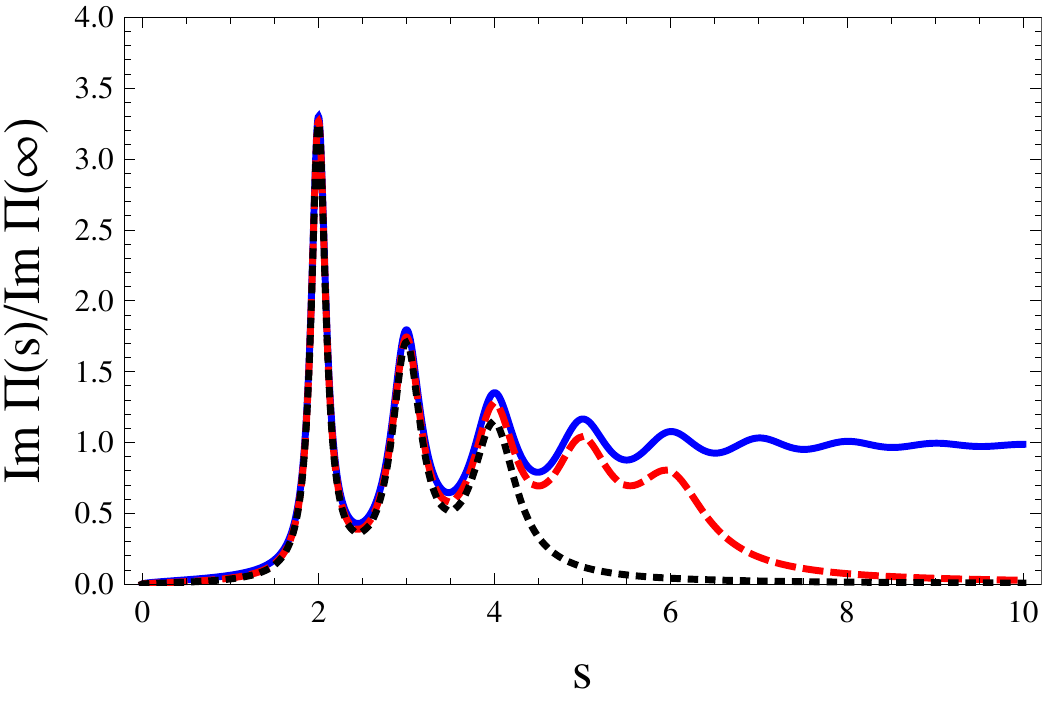}
\end{center}
\vspace{-3mm}
\caption{%Quark-hadron duality at work. 
Asymptotically normalized absortive part for the baryon polarization operator as a function of the CM squared energy when $M_n^2= n + 1$ and $ M_n \Gamma_n = \gamma n $ for $n=0,\dots,N$. Left panel: $\gamma=0.05$ (solid), $0.10$ (dashed), $0.15$ (dotted) for $N=100$.  
Right panel: $N=10$ (solid), $5$ (dashed) , $3$ (dotted) for
$\gamma=0.1$.
\label{fig:meson-regge}}
\end{figure} 

\section{Finite widths effects}

A remarkable feature of hadronic resonances noticed by Suranyi as
early as 1967~\cite{Domokos:1971aw} is that they are narrow, since
$\gamma \equiv \Gamma /M \sim 0.13$, a natural value in the light of
the large-$N_c$ expansion, where $\gamma= {\cal O}
(N_c^{-1})$~\cite{Witten:1979kh}. An upgrade of the Suranyi ratio
yields in average the value $\Gamma/M = 0.12(8)$ {\it both} for mesons
and baryons~\cite{Arriola:2011en}. This has interesting implications
for QHD at {\it finite} energies. For the correlator~(\ref{eq:corr}),
the finite width of the resonances can be implemented in
Eq.~(\ref{eq:dirac}) by an energy-dependent Breit-Wigner
distribution,
\begin{eqnarray}
\frac1{\pi} {\rm Im} \Pi(s) =  \frac1{\pi} \sum_{n=0}^N 
\frac{f_n^2  \Gamma_n \sqrt{s}}{(s-M_n^2)^2 + \Gamma_n^2 s} \, , 
\end{eqnarray}
where we put a cut-off $N$ in the sum and assume no quark
thresholds. We make the following {\it ansatz} compatible with
Eq.~(\ref{eq:qhd}) and $\Gamma_n / M_n \to \gamma$:
\begin{eqnarray}
f_n^2 = 1/(24 \pi^2 N_c) \, , \quad M_n^2 = n+1 \, , \quad M_n \Gamma_n = \gamma n \, , \quad n=0,1,2, \dots , N  \, . 
\end{eqnarray}
The dependence of ${\rm Im} \Pi(s)$ on both Suranyi's ratio $\gamma$
and the high energy cut-off $N$ is illustrated in
Fig.~\ref{fig:meson-regge}. As we see, the number of visible peaks
depends strongly on this ratio, but regardless of its numerical value
we find the {\it finite} limit at large $s$ which not only corresponds
to the narrow resonance limit, but also coincides with the finite $s$
result averaged over resonances.  
%~\footnote{This can be checked
%analytically since for large $s$ we can take the narrow resonance
%limit where the Breit-Wigner form becomes a Dirac-delta and replace
%the summation by an integral.}. 
On the other hand, if we cut-off the sum over states, we comply with
the expected partonic behaviour {\it only} below the cut-off,
i.e. $s \lesssim M_N^2 \sim N M_0^2$.  Similar patterns hold for the
correlator in Eq.~(\ref{eq:ncorr}) for finite $Q^2$ and
$x$~\cite{Domokos:1971aw}.

\begin{figure}[b]
\begin{center}
\includegraphics[width=0.7\textwidth]{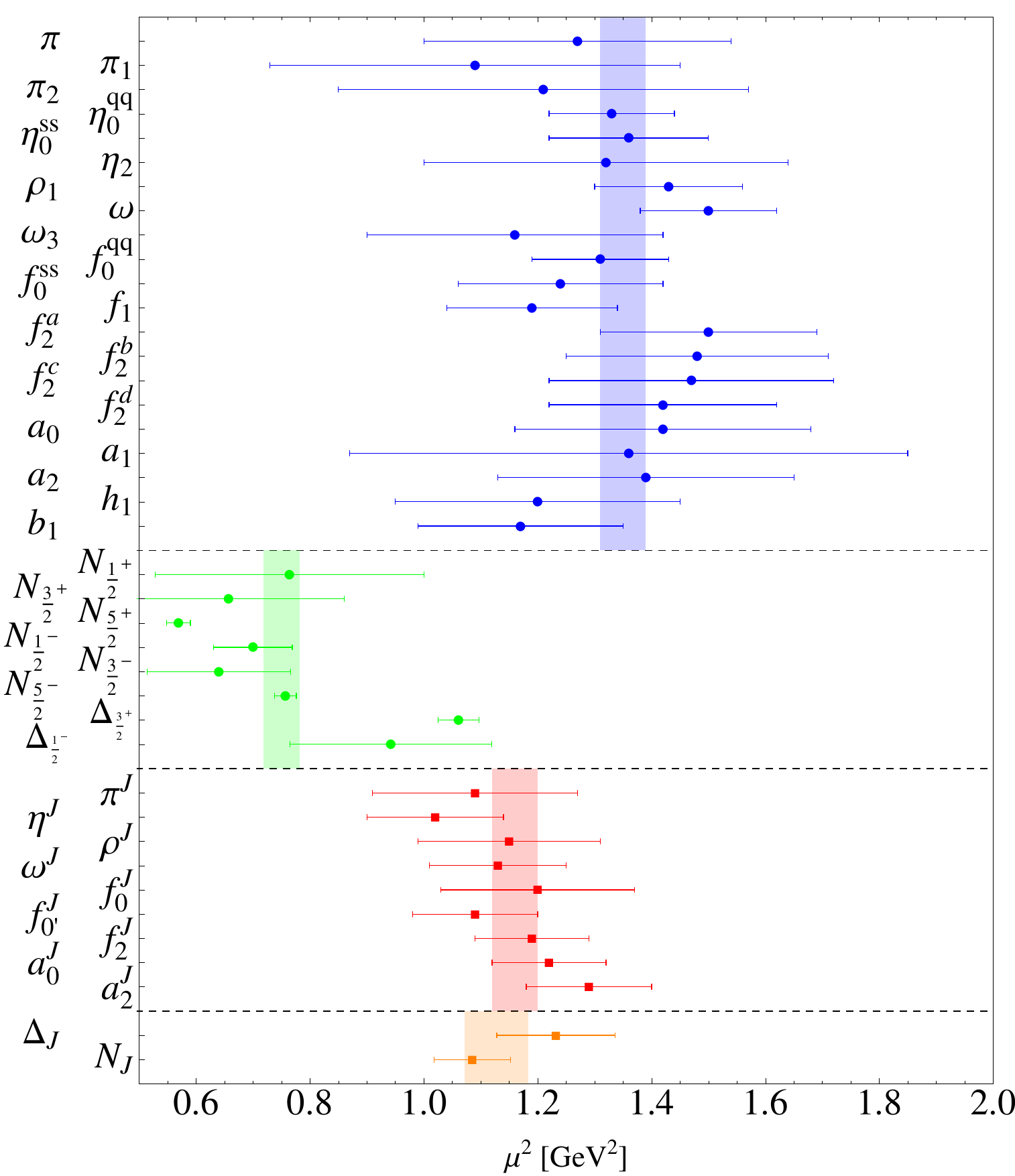}  
\end{center}
\vspace{-3mm}
\caption{Radial and angular slopes for non-strange mesons or baryons containing $(u,d)$ quarks and anti-quarks or diquarks respectively. 
\label{fig:baryon-meson-regge}}
\end{figure}

\section{Two body dynamics with linear potential and Regge fits}

For highly excited states relativity can be accomodated by the 
Salpeter equation~\cite{Arriola:2006sv}, where the (classical) mass
operator in the CM frame reads
\begin{eqnarray}
M = \sqrt{p^2 + m_1^2} + \sqrt{p^2 + m_2^2} + V_{12} (r) \, . 
\end{eqnarray} 
The spinor structure is included in the potential $V_{12} (r)$,
which accounts for the static interaction in the heavy quark limit,
$m_Q,m_{\bar Q} \gg p$, implying $V_{\bar Q Q} (r) \to\sigma_{\bar Q
Q} r$ for $\bar Q Q $ states (from charmonium data $\sigma_{\bar QQ} \sim
4.5 {\rm fm}^{-2}$~\cite{Segovia:2011tb}). For light quarks
$m_q,m_{\bar q} \to 0$ and  $M \to 2 p
+ \sigma_{\bar q q} r $ with $p^2= p_r^2 + L^2/r^2$. At large $M$,
we take $p \to p_r$, and the semiclassical quantization rule
$ \oint p_r dr = 2 \pi (n + \alpha)$, with $\alpha $ denoting a constant, leads
to~\cite{Arriola:2006sv}
% The turning points are $r=0$ and $r=M/\sigma$ so
%that
%\begin{eqnarray}
%2 \int_0^{M/\sigma} | M- \sigma r| dr = 2 \pi (n
%+ \alpha) = \frac{M^2}{2\sigma}  \to 
\begin{eqnarray}
 M_n^2 = \mu^2 n + M_0^2 \, , 
\label{eq:Mn}
\end{eqnarray}
with $\mu_{\bar q q}^2 = 4 \pi \sigma_{\bar q q} $. This agrees with
QHD of Eq.~(\ref{eq:qhd}) if $f_n^2 \to \mu^2 /( 24 \pi^2 N_c)$.  From the QHD
condition for baryons, Eq.~(\ref{eq:qhd-bar}), we get a mass formula
similar to Eq.~(\ref{eq:Mn}) and we infer a quark-diquark (qD)
long-distance dynamics with $V_{q D} (r) \to \sigma_{qD} r $ and $\mu_{qD}^2 =
4 \pi \sigma_{qD}$ of the form proposed in
Ref.~\cite{DeSanctis:2014ria} for the non-strange sector; their
$\sigma_{qD}=1.57 {\rm fm}^{-2}$ (called $\beta$ there) produces
$\mu_{qD}^2 = 0.76 {\rm GeV}^2$.
%Remarkably, the value for
%$\Sigma= \mu_{qD}^2/(4\pi)= $ matches a recent relativistic
%quark-diquark model analysis
%\section{Regge fits}
The QHD scaling requirements at the hadronic level, Eqs.~(\ref{eq:qhd})
and (\ref{eq:qhd-bar}), are consistent with an equidistant mass squared
spectrum for the intermediate hadronic states. More generally, Regge
trajectories can be parameterized as
%\begin{equation} 
$M_{n,J}^2 = a + \mu^2 n + \beta^2 J $. 
%\end{equation}
Our fits for the radial and angular slopes in the case of non-strange
mesons~\cite{Masjuan:2012gc} and baryons~\cite{Masjuan:2017fzu} assume 
an uncertainty given by the finite width of the resonances:
\begin{eqnarray}
\chi^2 = \sum_\alpha \left(\frac{M_\alpha^2 - (M_{\alpha}^{\rm exp})^2}{\Gamma^{\rm exp}_\alpha M^{\rm exp}_\alpha}\right)^2 \, ,  
\end{eqnarray}
and are summarized in Fig.~\ref{fig:baryon-meson-regge}.  We find that
$\mu_{q \bar q}^2 = 1.34(4) {\rm GeV}^2 \sim 2 \mu_{q D}^2 $ but
$\beta_{q \bar q}^2 \sim \beta_{qD}^2 \sim 1.15 {\rm GeV}^2$. The
value $\mu_{qD}^2 = 0.75(3) {\rm GeV}^2$ agrees remarkably well with
the relativistic qD model findings~\cite{DeSanctis:2014ria}. Thus, our
QHD-based analysis supports the idea that possibly $ {\cal H}_{\rm PDG}
= {\cal H}_{\bar q q} \oplus {\cal H}_{qD} \oplus {\cal H}_{\bar
q \bar D} \oplus \dots $.

%\bibliographystyle{h-elsevier}
%\bibliography{../Paper/v2/diquarks,excitedQCD}

\end{document}